# Moiré Superlattice Modulations in Single-Unit-Cell FeTe Films Grown on NbSe$_2$ Single Crystals

Han-Bin Deng (邓翰宾)[1,2,+], Yuan Li (李渊)[1,2,+], Zili Feng (冯子力)[1,2], Jian-Yu Guan (关剑宇)[1,2], Xin Yu (于鑫)[1], Xiong Huang (黄雄)[1,2], Rui-Zhe Liu (刘睿哲)[1,2], Chang-Jiang Zhu (朱长江)[1,2], Limin Liu (刘立民)[1,2], Ying-Kai Sun (孙英开)[1,2], Xi-liang Peng (彭锡亮)[1,2], Shuai-Shuai Li (李帅帅)[1,2], Xin Du (杜鑫)[1,2], Zheng Wang (王铮)[1,2], Rui Wu (武睿)[1,3], Jia-Xin Yin (殷嘉鑫)[4], You-Guo Shi (石友国)[1,3,5] and Han-Qing Mao (毛寒青)[1,*]

[1]*Beijing National Laboratory for Condensed Matter Physics and Institute of Physics, Chinese Academy of Sciences, Beijing 100190, China.*
[2]*School of Physics, University of Chinese Academy of Sciences, Beijing 100049, China.*
[3]*Songshan Lake Materials Laboratory, Dongguan, Guangdong 523808, China.*
[4]*Laboratory for Topological Quantum Matter and Spectroscopy (B7), Department of Physics, Princeton University, Princeton, NJ 08544, USA.*
[5]*Center of Materials Science and Optoelectronics Engineering, University of Chinese Academy of Sciences, Beijing 100049, China.*

## ABSTRACT

Interface can be a fertile ground for exotic quantum states, including topological superconductivity, Majorana mode, fractal quantum Hall effect, unconventional superconductivity, Mott insulator, etc. Here we grow single-unit-cell (1UC) FeTe film on NbSe$_2$ single crystal by molecular beam epitaxy (MBE) and investigate the film *in-situ* with home-made cryogenic scanning tunneling microscopy (STM) and non-contact atomic force microscopy (AFM) combined system. We find different stripe-like superlattice modulations on grown FeTe film with different misorientation angles with respect to NbSe$_2$ substrate. We show that these stripe-like superlattice modulations can be understood as moiré pattern forming between FeTe film and NbSe$_2$ substrate. Our results indicate that the interface between FeTe and NbSe$_2$ is atomically sharp. By STM-AFM combined measurement, we suggest the moiré superlattice modulations have an electronic origin when the misorientation angle is relatively small (≤3°) and have structural relaxation when the misorientation angle is relatively large (≥10°).

[+]These authors contributed equally to this work.
[*]Corresponding author: mhq@iphy.ac.cn.

# 1. INTRODUCTION

Interface engineering may lead to many exotic quantum phenomena. In recent years, topological superconductivity and Majorana fermions for fault-tolerant quantum computation have been found in fabricated interface systems with a topological insulator and a conventional s-wave superconductor.[1-8] On the other hand, fractal quantum Hall effect, unconventional superconductivity, and Mott insulator have been found in fabricated interface system with moiré superlattice which significantly modifies its electronic properties.[9-14]

Non-trivial topology has been proposed in Fe(Te,Se) single crystal and thin film.[15-18] While *2H*-NbSe$_2$ single crystal is a prototype of conventional s-wave superconductor. If the two materials are used to fabricate the interface system, non-trivial topology and superconductivity may take place.[1,2] However, the two materials are in different crystallographic symmetries. Fabricating an atomically sharp interface between them is not an easy task and misorientation of different angles between them is expected. In addition, the different heights of Te atom and Se atom may hinder the STM/AFM investigation of the interface structure between Fe(Te,Se) film and NbSe$_2$ substrate. Therefore, FeTe film which has the same lattice structure but a simpler chemical composition may be an alternate. According to the DFT calculations of X. X. Wu,[17] 1UC FeTe becomes topological non-trivial when $a_{\mathrm{FeTe}} < 3.905$Å. Although the band topology of 1UC FeTe film grown on NbSe$_2$ ($a_{\mathrm{FeTe}} \approx 3.9$Å) is still not clear, it is wise to use 1UC FeTe/NbSe$_2$ as a prototype of 1UC Fe(Te,Se)/NbSe$_2$ to investigate their interface structure. Therefore, the influence of misorientation on the structural relaxation and/or electronic reconstruction of FeTe/NbSe$_2$ may be explored.

Here in this paper, we grow atomically flat 1UC FeTe film on *2H*-NbSe$_2$ single crystal by MBE system and investigate it *in-situ* with home-built STM/AFM. In the STM topographic image of grown FeTe film, we find different stripe-like superlattice modulations associated with different misorientation angles with respect to NbSe$_2$ substrate. By superimposing the tetragonal FeTe lattice on the hexagonal NbSe$_2$ lattice, we reproduce the key features of the stripe-like superlattice modulations. Therefore, our results show that the stripe-like superlattice modulations are moiré superlattice modulations, indicating an atomically sharp interface between FeTe film and NbSe2 substrate. Our results also suggest an electronic origin of the moiré superlattice modulations in most of the cases when the misorientation angle is relatively small (≤3°), because we find the moiré superlattice modulations are absent in AFM topographic image. While when the misorientation angle is relatively large (≥10°), structural relaxation emerges with the larger periodicity of the moiré superlattice modulations.

# 2. METHODS

The *2H*-NbSe$_2$ (0001) substrates typically 2 × 2 × 0.5 mm in size are glued on a titanium holder with H20E from *Epoxy Technology*. The substrates are transferred into our home-built cryogenic SPM-MBE combined system.[19] Then the substrates are cleaved in a vacuum after degassing at 320 °C for at least 6 hours. So large, flat, and clean surfaces are achieved.

1UC FeTe (001) films are epitaxially grown on substrates with co-evaporation of high-purity Fe

(99.98%) and Te (99.99%) from standard Knudsen cells for 6~12 minutes, while the substrates are heated to 260 °C. The stable fluxes of Fe and Te sources are obtained by heating the cells to 1045 °C and 300 °C, respectively. The high-quality films are achieved by post-annealing at 260 °C for ~4hs. After post-annealing, the FeTe films are quickly transferred to an *in-situ* cryogenic STM/AFM for measurement.

Both STM and AFM measurements can be applied by our home-made cryogenic STM/AFM system. When using an STM tip holder, the traditional STM function can be realized. When using an AFM tip holder of qPlus configuration, amplitude modulated (AM) or frequency modulated (FM) AFM measurement can be realized. By an additional electrode connecting the metal tip of the qPlus sensor, tunneling current can be picked up during AFM measurement, simultaneously. Electrochemically etched tungsten tips are used in both STM and AFM mode, which are *in-situ* modified by gently touching Ag(111) film surface. All STM topographic images are obtained in constant current mode at 4K. The typical Q factor of the qPlus sensor is ~ 35000 at 4K, with resonant frequency at about 25kHz.

The step height of 1UC FeTe film is ~7.4 Å, relatively larger than the step height of the second-unit-cell FeTe film, which is ~6.3 Å.

The misorientation angle $\varphi$ is measured by comparing FeTe lattice vectors with respect to NbSe$_2$ lattice vectors. Only one domain has been found in each NbSe$_2$ substrate. So the lattice vectors of NbSe$_2$ can either be measured before or after film deposition.

## 3. RESULTS and DISCUSSION

*2H*-NbSe$_2$ and FeTe are layered materials with different crystallographic symmetries, as shown in FIG. 1(a). The unit cell of *2H*-NbSe$_2$ consists of two hexagonal Nb layers. Each Nb layer is sandwiched by two hexagonal Se layers. The weakest bonding is the Van der Waals (VdW) interaction between adjacent Se layers, where cleavage takes place. While the unit cell of FeTe consists of one tetragonal Fe layer sandwiched by two tetragonal Te layers. The bonding between adjacent Te layers is also VdW interaction. Therefore, the bonding between FeTe film and NbSe$_2$ substrate is also expected to be weak. And misorientation of different angles may take place. As shown in FIG. 1(b), the misorientation angle $\varphi$ is defined as the minimum angle between FeTe lattice vectors ($a_{\text{Te−Te}}$) with respect to NbSe$_2$ lattice vectors ($a_{\text{Se−Se}}$). Most of the grains of the grown FeTe films have relatively small misorientation angles ($\varphi \leq 3°$). However, grains with misorientation angles between 3° to 15° have also been found. FIG. 1(c) and FIG. 1(e) show two typical atomic-resolved STM topographic images in 1UC FeTe film, with the misorientation angle $\varphi$ ~0° and ~15°, respectively. The lattice periodicity of the topmost Te layer is measured ~3.9 Å by the reciprocal vectors in their corresponding FFT images (indicated by the red arrows in FIG. 1(d) and FIG. 1(f)). Besides the lattice periodicity, nanoscale stripe-like superlattice modulations can also be resolved. Comparing FIG. 1(c) with FIG. 1(e), we find the stripe-like superlattice modulations are different when the misorientation angle varies. FIG. 1(d) and FIG. 1(f) are their corresponding FFT images. When the misorientation angle $\varphi$ is ~0°, three pairs of wave vectors (indicated by the red circles in FIG. 1(d)) show similarly strongest intensity. However, only one pair of wave vectors (indicated by the red circles in FIG. 1(f)) shows the strongest intensity when the misorientation angle $\varphi$ is ~15°. The variation of the stripe-like superlattice modulations when the misorientation angle $\varphi$ changes can also be found near the grain boundaries. FIG. 1(g) is an atomic-resolved STM topographic image probed near a domain boundary. In its FFT image FIG. 1(h), two groups of reciprocal vectors can be resolved, as indicated by the two red arrows. We find that the misorientation angle $\varphi$ is ~1.5° in the bottom-right domain and ~10.5° in the top-left domain. In the bottom-right domain when $\varphi$ is small, the stripe-like superlattice modulations resemble the ones in FIG. 1(c). While in the top-left domain when $\varphi$ is relatively larger, the stripe-like superlattice modulations look similar to the ones in FIG. 1(e). The dependence of the stripe-like superlattice modulations on the misorientation angle $\varphi$ is consistent in our different samples, implying it is the intrinsic property of 1UC FeTe/NbSe$_2$ film.

Several previous reports have found stripe-like moiré superlattice modulations in FeSe/Bi$_2$Se$_3$ and FeSe/Pb interface systems.[20-25] Therefore, the stripe-like superlattice modulations found in FeTe/NbSe$_2$ film may also form by the moiré pattern between film and substrate. By superimposing the tetragonal lattice of FeTe film (red circles) on the hexagonal lattice of NbSe$_2$ substrate (blue circles, here the CDW modulation in NbSe$_2$ is excluded for simplicity), we simulate the moiré pattern between epitaxial FeTe films and NbSe$_2$ substrates.[20,26] The simulated moiré patterns are shown in FIG. 2(a), FIG. 2(b) and FIG. 2(c), respectively. Comparing them with the measured STM topographic images, we find the simulated moiré patterns reproduce the periodicities and the directions of the stripe-like superlattice modulations in all the cases. Hence our simulation strongly supports that the stripe-like superlattice modulations found in 1UC FeTe/NbSe$_2$ films are also moiré superlattice modulations, which are driven by the influence of the crystal field of the NbSe$_2$ substrate on FeTe film.

The periodic crystal field of the NbSe$_2$ substrate may lead to structural relaxation or electronic reconstruction on the FeTe film. Since the non-contact AFM topographic image is less sensitive to the electronic structures than the STM topographic image,[27-29] it is beneficial to recheck the moiré superlattice modulations with AFM measurement. Most of the FeTe grains have relatively small misorientation angles ($\varphi \leq 3°$) and their properties are similar. FIG. 3(a) is an atomic-resolved STM topographic image of FeTe film when the misorientation angle $\varphi$ is ~3°. FIG. 3(b) is an AFM topographic image of the same field of view (FOV), in which the moiré superlattice modulations can not be resolved. While in the simultaneously obtained current image from the tungsten tip, as shown in FIG. 3(c), the moiré superlattice modulations can still be resolved. In relatively rare cases, FeTe grains with larger angles can also be found. FIG. 3(d) and (g) are atomic-resolved STM topographic images of FeTe film when the misorientation angles are ~7° and ~12°, respectively. FIG. 3(e) and (h) are atomic-resolved AFM topographic images of the same FOVs, respectively. When the misorientation angle $\varphi$ is ~7°, the moiré pattern in the AFM topological image can hardly be resolved. While when the misorientation angle $\varphi$ is ~12°, the moiré pattern in the AFM topological image can be clearly resolved. In the simultaneously obtained current image from the tungsten tip, as shown in FIG. 3(f) and (i), the moiré superlattice modulations can also be resolved. Hence our STM-AFM combined results suggest an electronic origin of the moiré superlattice modulations when the misorientation angle $\varphi$ is relatively small ($\leq 3°$) and structural relaxation emerges when the misorientation angle $\varphi$ is relatively small ($\geq 10°$).

FIG. 4(a) plots the measured corrugation of the moiré superlattice modulations in the AFM topographic image with respect to $\varphi$. When the misorientation angle $\varphi$ is relatively small ($\leq 3°$), the moiré pattern can not be resolved in the AFM topographic image and the corrugation is 0 pm. When the misorientation angle $\varphi$ increases and is larger than 10°, a sharp enhancement of the moiré pattern corrugation above 20 pm can be found. FIG. 4(b) plots the measured periodicities of the strongest moiré pattern wave vectors from the FFT of the STM topographic images. When the misorientation angle $\varphi$ is relatively small ($\leq 3°$), the periodicities of the three pairs of the moiré pattern wave vectors are similarly short. When the misorientation angle $\varphi$ increases and is larger than 10°, only one pair of the moiré pattern wave vectors has the strongest intensity and a sharp increase of its periodicity is found. A similar relationship between structural relaxation and misorientation angle $\varphi$ has been reported in layered material $MoS_2$[30]. The increased periodicities of the moiré superlattice modulations increase the distance between neighboring Te-Se atom stacking, allowing lattice relaxation even though it is driven by weak VdW interaction[30].

## 4. CONCLUSIONS

In summary, we successfully grow and investigate 1UC FeTe film on $NbSe_2$ single crystal with a homemade MBE-STM/AFM combined system. We find different stripe-like superlattice modulations in the STM topographic images of grown FeTe film associated with its different misorientation angles with respect to NbSe2 substrate. Comparing the experimental results with simulation by superimposing FeTe and $NbSe_2$, we unambiguously show the stripe-like superlattice modulations are moiré superlattice modulations in different misorientation angles. Therefore, our results support the interface between FeTe film and $NbSe_2$ substrate is atomically sharp.

Furthermore, we find the moiré superlattice modulations have an electronic origin with STM-AFM combined experiments in most cases when the misorientation angle $\varphi$ is relatively small ($\leq 3°$). In the FM AFM topographic image, the moiré superlattice modulations are hardly distinguished, but still clearly resolved in the simultaneously obtained current image. Since STM topographic image is more sensitive to electronic structure than AFM topographic image, we suggest that the moiré patterns are modulations of the local density of states other than structural relaxation. However, in rare cases when the misorientation angle $\varphi$ is relatively large ($\geq 10°$), structural relaxation emergies and can be resolved in the AFM topographic image. Therefore, the interface system between FeTe film and $NbSe_2$ substrate may be a platform to study the influence of misorientation on the structural relaxation and/or electronic reconstruction of $FeTe/NbSe_2$.

## ACKNOWLEDGEMENT

*Project supported by the National Key Research and Development Program of China (Grants No. 2016YFA0302400, No. 2016YFA0300602, and No. 2017YFA0302903), the National Natural Science Foundation of China (Grant No. 11227903), the Beijing Municipal Science and Technology Commission (Grants No. Z181100004218007 and No. Z191100007219011), the National Basic Research Program of China (Grant No. 2015CB921304), and the Strategic Priority Research Program of Chinese Academy of Sciences ( Grants No. XDB07000000, No. XDB28000000, and No. XDB33000000).

# FIGURES and TABLES

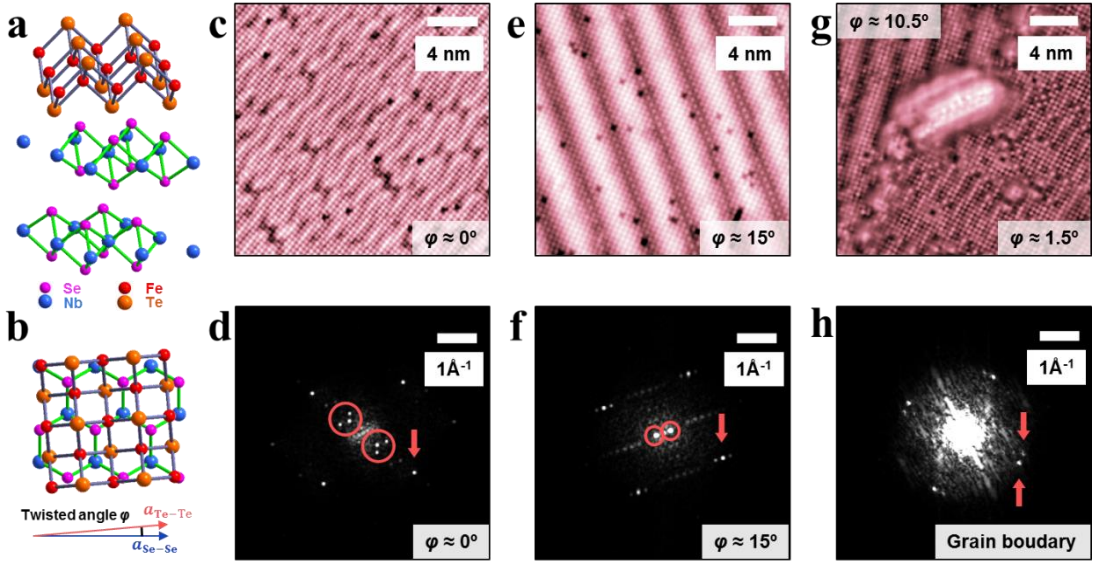

FIG. 1. (a) Illustration of 1UC FeTe film grown on NbSe$_2$ single crystal. (b) Top view of 1UC FeTe film grown on NbSe$_2$ single crystal. The misorientation angle $\varphi$ is defined as the minimum angle of FeTe lattice vectors ($a_{\text{Te}-\text{Te}}$) with respect to NbSe$_2$ lattice vectors ($a_{\text{Se}-\text{Se}}$). Only topmost Se-Nb-Se layers are plotted for simplicity. (c) An atomic-resolved STM image of 1UC FeTe film when the misorientation angle $\varphi$ is ~0°. The junction setup is: $V_{\text{bias}}$ = 100mV, $I_{\text{set}}$ = 1.8nA. (d) The FFT image of (c). The reciprocal vectors of the tetragonal Te layer and strongest superlattice modulations are marked by the red arrow and the red circles, respectively. The periodicities of the three pairs of strongest superlattice modulations are ~13.6 Å, ~10.4 Å, and ~10.2 Å, respectively. (e) An atomic-resolved STM image of 1UC FeTe film when the misorientation angle $\varphi$ is ~15° (the max value of misorientation angle between a tetragonal lattice and a hexagonal lattice). The junction setup is: $V_{\text{bias}}$ = 100mV, $I_{\text{set}}$ = 100pA. (f) The FFT image of (e). The reciprocal vectors of the tetragonal Te layer and strongest stripe-like superlattice are marked by the red arrow and the red circles, respectively. The periodicity of the pair of strongest superlattice modulation is ~31.7 Å. (g) An atomic-resolved STM image of 1UC FeTe film near a domain boundary. The misorientation angles $\varphi$ on two sides of the grain boundary are ~1.5° and ~10.5°, respectively. The junction setup is: $V_{\text{bias}}$ = 10mV, $I_{\text{set}}$ = 200pA. (h) The FFT image of (g). The reciprocal vectors of tetragonal Te lattice on two sides of the grain boundaries are marked by the two red arrows.

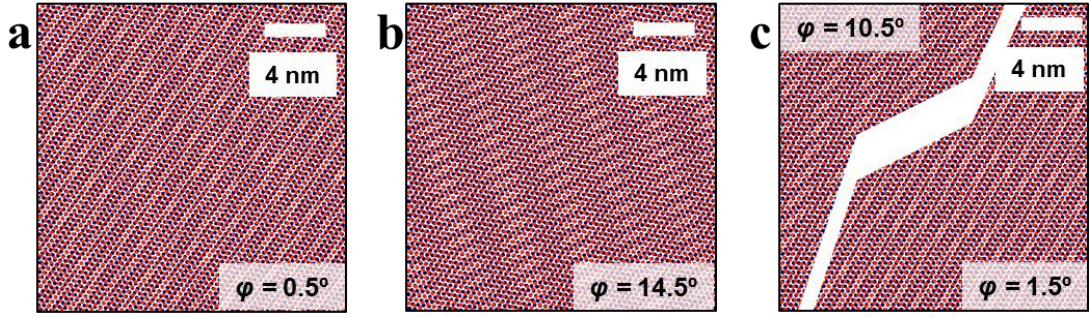

FIG. 2. (a) The simulated moiré pattern between FeTe lattice and NbSe$_2$ lattice in FIG. 1(c). The misorientation angle $\varphi$ is set to 0.5°. (b) The simulated moiré pattern between FeTe lattice and NbSe$_2$ lattice in FIG. 1(e). The misorientation angle $\varphi$ is set to 14.5°. (c) The simulated moiré pattern between FeTe lattice and NbSe$_2$ lattice near a grain boundary in FIG. 1(g). The misorientation angle $\varphi$ in the bottom-right FeTe domain is set to 1.5°, and the misorientation angle $\varphi$ in the top-left FeTe domain is set to 10.5°. The grain boundary is indicated by the thick white polyline in the graph.

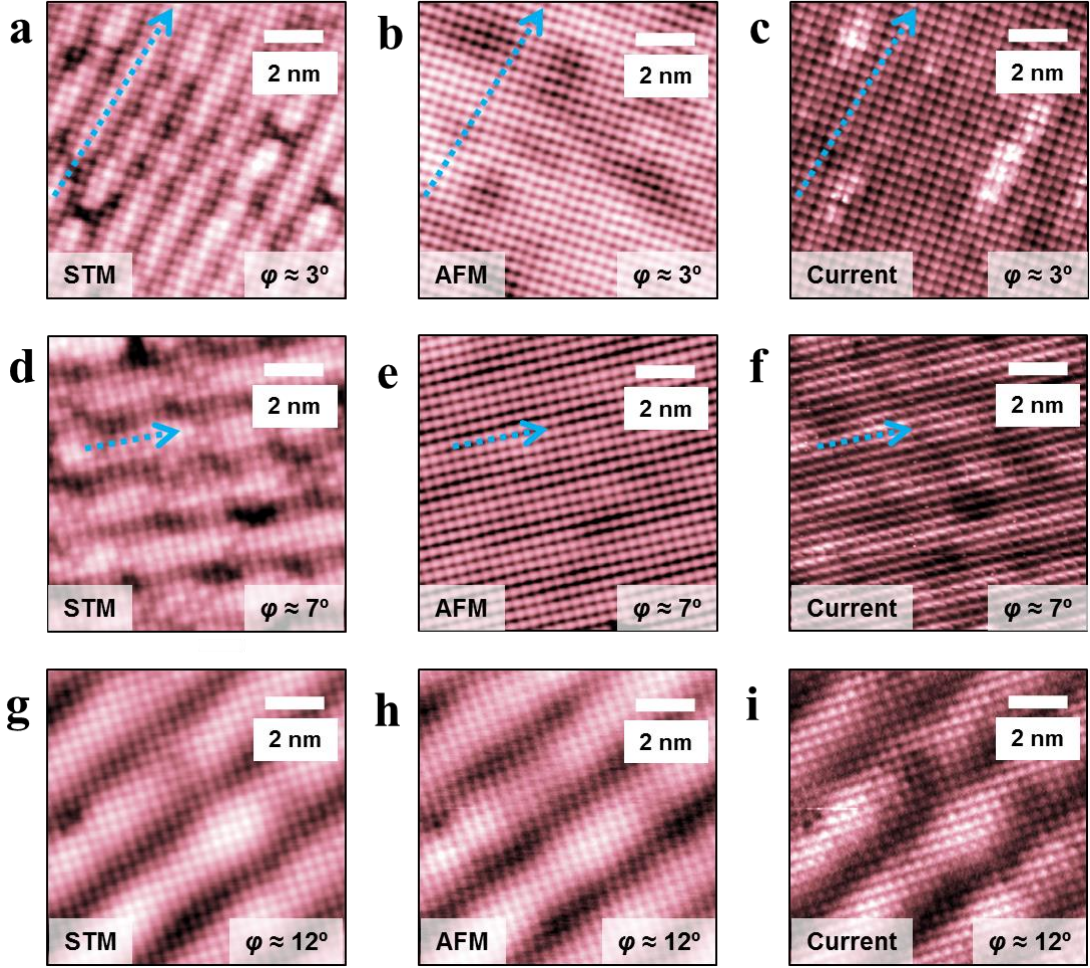

FIG. 3. (a) STM topographic images (the misorientation angle is ~3°) obtained by a qPlus sensor in a constant current mode, while AFM signals are grounded. The junction setup is: $V_{bias}$ = 100mV, $I_{set}$ = 100pA. (b) AFM topographic images of the same FOV to (a) obtained by a qPlus sensor in a constant frequency shift mode with Δf = -6.5Hz and $V_{bias}$ = 10mV. The oscillation frequency of the quartz tuning fork is about 22.6kHz, which is far larger than the cut-off frequency of the STM preamplifier (~1.1kHz). A small bias voltage of 10mV is applied to the sample, so the averaged current signal can be obtained simultaneously. The modulations of the moiré pattern can not be resolved. (c) Current image obtained from the tip simultaneously with (b). The modulations of the moiré pattern can be resolved. The blue arrows in (a-c) indicate the direction of the stripe-like moiré superlattice modulations in (a). (d) STM topographic images when the misorientation angle is ~7°. The junction setup is: $V_{bias}$ = 100mV, $I_{set}$ = 10pA. (e) AFM topographic images of the same FOV to (d) with Δf = -7Hz and $V_{bias}$ = 10mV. The oscillation frequency of the quartz tuning fork is about 32.2kHz. The modulations of the moiré pattern can hardly be resolved. (f) Current image obtained from the tip simultaneously with (e). The modulations of the moiré pattern can be resolved. The blue arrows in (d-f) indicate the direction of the stripe-like moiré superlattice modulations in (d). (g) STM topographic images when the misorientation angle is ~12°. The junction setup is: $V_{bias}$ = 10mV, $I_{set}$ = 100pA. (h) AFM topographic images of the same FOV to (g) with Δf = -8.8Hz and $V_{bias}$ = 10mV. The oscillation frequency of the quartz tuning fork is about 32.2kHz. The modulations of the moiré pattern can be well resolved. (i) Current image obtained from the tip simultaneously with (h). The modulations of the moiré pattern can also be resolved.

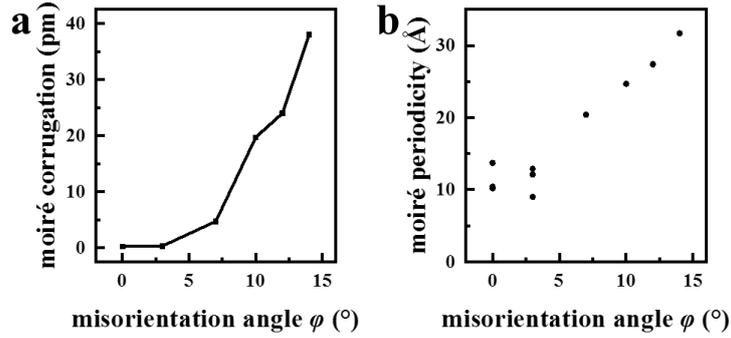

FIG. 4. (a) The measured corrugation of the moiré pattern in the AFM topographic image with respect to $\varphi$. (b) The measured periodicities of the moiré superlattice modulations in the FFT of the STM topographic image with respect to $\varphi$.

# REFERENCES

1   Fu L and Kane C L 2008 *Phys. Rev. Lett.* **100** 096407
2   Linder J, Tanaka Y, Yokoyama T, Sudbø A and Nagaosa N 2010 *Phys. Rev. Lett.* **104** 067001
3   Nayak C, Simon S H, Stern A, Freedman M and Das Sarma S 2008 *Rev. Mod. Phys.* **80** 1083
4   Wang M-X, Liu C, Xu J-P, Yang F, Miao L, Yao M-Y, Gao C L, Shen C, Ma X, Chen X, Xu Z-A, Liu Y, Zhang S-C, Qian D, Jia J-F and Xue Q-K 2012 *Science* **336** 52
5   Xu J-P, Liu C, Wang M-X, Ge J, Liu Z-L, Yang X, Chen Y, Liu Y, Xu Z-A, Gao C-L, Qian D, Zhang F-C and Jia J-F 2014 *Phys. Rev. Lett.* **112** 217001
6   Xu S-Y, Alidoust N, Belopolski I, Richardella A, Liu C, Neupane M, Bian G, Huang S-H, Sankar R, Fang C, Dellabetta B, Dai W, Li Q, Gilbert M J, Chou F, Samarth N and Hasan M Z 2014 *Nat. Phys.* **10** 943
7   Xu J-P, Wang M-X, Liu Z L, Ge J-F, Yang X, Liu C, Xu Z A, Guan D, Gao C L, Qian D, Liu Y, Wang Q-H, Zhang F-C, Xue Q-K and Jia J-F 2015 *Phys. Rev. Lett.* **114** 017001
8   Sun H-H, Zhang K-W, Hu L-H, Li C, Wang G-Y, Ma H-Y, Xu Z-A, Gao C-L, Guan D-D, Li Y-Y, Liu C, Qian D, Zhou Y, Fu L, Li S-C, Zhang F-C and Jia J-F 2016 *Phys. Rev. Lett.* **116** 257003
9   Dean C R, Wang L, Maher P, Forsythe C, Ghahari F, Gao Y, Katoch J, Ishigami M, Moon P, Koshino M, Taniguchi T, Watanabe K, Shepard K L, Hone J and Kim P 2013 *Nature* **497** 598
10  Cao Y, Fatemi V, Fang S, Watanabe K, Taniguchi T, Kaxiras E and Jarillo-Herrero P 2018 *Nature* **556** 43
11  Chen G, Sharpe A L, Gallagher P, Rosen I T, Fox E J, Jiang L, Lyu B, Li H, Watanabe K, Taniguchi T, Jung J, Shi Z, Goldhaber-Gordon D, Zhang Y and Wang F 2019 *Nature* **572** 215
12  Lu X, Stepanov P, Yang W, Xie M, Aamir M A, Das I, Urgell C, Watanabe K, Taniguchi T, Zhang G, Bachtold A, MacDonald A H and Efetov D K 2019 *Nature* **574** 653
13  Cao Y, Fatemi V, Demir A, Fang S, Tomarken S L, Luo J Y, Sanchez-Yamagishi J D, Watanabe K, Taniguchi T, Kaxiras E, Ashoori R C and Jarillo-Herrero P 2018 *Nature* **556** 80
14  Chen G, Jiang L, Wu S, Lyu B, Li H, Chittari B L, Watanabe K, Taniguchi T, Shi Z, Jung J, Zhang Y and Wang F 2019 *Nat. Phys.* **15** 237
15  Yin J X, Wu Z, Wang J H, Ye Z Y, Gong J, Hou X Y, Shan L, Li A, Liang X J, Wu X X, Li J, Ting C S, Wang Z Q, Hu J P, Hor P H, Ding H and Pan S H 2015 *Nat. Phys.* **11** 543
16  Peng X L, Li Y, Wu X X, Deng H B, Shi X, Fan W H, Li M, Huang Y B, Qian T, Richard P, Hu J P, Pan S H, Mao H Q, Sun Y J and Ding H 2019 *Phys. Rev. B* **100** 155134
17  Wu X, Qin S, Liang Y, Fan H and Hu J 2016 *Phys. Rev. B* **93** 115129
18  Shi X, Han Z-Q, Richard P, Wu X-X, Peng X-L, Qian T, Wang S-C, Hu J-P, Sun Y-J and Ding H 2017 *Sci. Bull.* **62** 503
19  Mao H Q *unpublished*
20  Wang Y, Jiang Y, Chen M, Li Z, Song C, Wang L, He K, Chen X, Ma X and Xue Q-K 2012 *J. Phys. Condens. Matter.* **24** 475604
21  Eich A, Rollfing N, Arnold F, Sanders C, Ewen P R, Bianchi M, Dendzik M, Michiardi M, Mi J-L, Bremholm M, Wegner D, Hofmann P and Khajetoorians A A 2016 *Phys. Rev. B* **94** 125437
22  Singh U R, Warmuth J, Markmann V, Wiebe J and Wiesendanger R 2016 *J. Phys. Condens.*

|    |    |
|----|----|
| | *Matter.* **29** 025004 |
| 23 | Cavallin A, Sevriuk V, Fischer K N, Manna S, Ouazi S, Ellguth M, Tusche C, Meyerheim H L, Sander D and Kirschner J 2016 *Surf. Sci.* **646** 72 |
| 24 | Song S Y, Martiny J H J, Kreisel A, Andersen B M and Seo J 2020 *Phys. Rev. Lett.* **124** 117001 |
| 25 | Qin H, Chen X, Guo B, Pan T, Zhang M, Xu B, Chen J, He H, Mei J, Chen W, Ye F and Wang G 2021 *Nano Lett.* **21** 1327 |
| 26 | Fu D, Zhao X, Zhang Y-Y, Li L, Xu H, Jang A R, Yoon S I, Song P, Poh S M, Ren T, Ding Z, Fu W, Shin T J, Shin H S, Pantelides S T, Zhou W and Loh K P 2017 *J. Am. Chem. Soc.* **139** 9392 |
| 27 | Hembacher S, Giessibl F J, Mannhart J and Quate C F 2003 *Proc. Natl. Acad. Sci.* **100** 12539 |
| 28 | Mao H-Q, Li N, Chen X and Xue Q-K 2012 *Chinese Phys. Lett.* **29** 066802 |
| 29 | Sun Z, Hämäläinen S K, Sainio J, Lahtinen J, Vanmaekelbergh D and Liljeroth P 2011 *Phys. Rev. B* **83** 081415 |
| 30 | Quan J, Linhart L, Lin M-L, Lee D, Zhu J, Wang C-Y, Hsu W-T, Choi J, Embley J, Young C, Taniguchi T, Watanabe K, Shih C-K, Lai K, MacDonald A H, Tan P-H, Libisch F and Li X 2021 *Nat. Mater.* |